\documentclass[letterpaper,twocolumn,10pt]{article}
\usepackage{usenix2019_v3}
\usepackage{enumerate}
\usepackage{silence}
\usepackage{amsmath,amsopn,amssymb}
\usepackage{subfigure}
\usepackage{endnotes,microtype,xspace,graphicx,fancyvrb,multirow}
\usepackage{booktabs}
\usepackage{array,underscore,relsize}
\usepackage[T1]{fontenc}
\usepackage{times}
\usepackage{fancyhdr,lastpage}
\usepackage{enumitem}
\usepackage[labelfont=bf,font=small,skip=1pt]{caption}
\usepackage{comment}
\usepackage{fp}
\usepackage{siunitx}

\newcommand{\sys}{\mbox{\textsc{Roki}}\xspace}

\newcommand{\cc}[1]{\mbox{\smaller[0.5]\texttt{#1}}}

\newcommand{\segmentsmack}{\mbox{SegmentSmack}\xspace}
\newcommand{\fragmentsmack}{\mbox{FragmentSmack}\xspace}
\newcommand{\udos}{\mbox{$\mu$DoS}\xspace}

\fvset{fontsize=\scriptsize,xleftmargin=8pt,numbers=left,numbersep=5pt}

\input{code/fmt}
\newcommand{\figrule}{\hrule width \hsize height .33pt}
\newcommand{\coderule}{\vspace{0.4em}\figrule\vspace{0.2em}}

\def\Snospace~{\S{}}

\newcommand{\x}{$\times$\xspace}

\newif\ifdraft\drafttrue
\newif\ifnotes\notestrue
\ifdraft\else\notesfalse\fi

\input{glyphtounicode}
\newcolumntype{R}[1]{>{\raggedleft\let\newline\\\arraybackslash\hspace{0pt}}p{#1}}

\newcommand{\includepdf}[1]{
  \includegraphics[width=\columnwidth]{#1}
}

\newcommand{\squishlist}{
\begin{itemize}[noitemsep,nolistsep]
  \setlength{\itemsep}{-0pt}
}
\newcommand{\squishend}{
  \end{itemize}
}

\usepackage{tikz}

\usepackage{xstring}
\newcommand{\PP}[1]{
\vspace{2px}
\noindent{\bf \IfEndWith{#1}{.}{#1}{#1.}}
}

\newcommand{\X}{{\footnotesize $\times$}\xspace}

\newcommand{\B}{\,\text{B}\xspace}

\newcommand{\KB}{\,\text{KB}\xspace}
\newcommand{\MB}{\,\text{MB}\xspace}

\newcommand{\ie}{i.e.}
\newcommand{\eg}{e.g.}
\newcommand{\naive}{na\"{i}ve\xspace}

\newcommand{\boxbeg}{
\vspace{4px}
\noindent
\vspace{2px}
\begin{tabular}{|l|}\hline
\begin{minipage}{0.94\columnwidth}
\vspace{2px}
\noindent
}

\newcommand{\boxend}{
\vspace{2px}
\end{minipage}\\ \hline
\end{tabular}
}

\gdef\therev{b4bba2a}
\gdef\thedate{2019-03-28 17:02:29 -0400}

\usepackage{titling}
\begin{document}

\title{\Large \bf Mitigating Low-volume DoS Attacks with Data-driven Resource Accounting}

\ifdefined\DRAFT
 \pagestyle{fancyplain}
 \lhead{Rev.~\therev}
 \rhead{\thedate}
 \cfoot{\thepage\ of \pageref{LastPage}}
\fi

\author{
ChangSeok Oh$^\dagger$\;
Sangho Lee$^\ast$\;
Wen Xu$^\dagger$\;
Rohan Devang Vora$^\dagger$\;
Taesoo Kim$^\dagger$\;
\\\\
\emph{$^\dagger$Georgia Institute of Technology}\;
\emph{$^\ast$Microsoft Research}
}


\date{}
\maketitle

\begin{abstract}
  Low-volume Denial-of-Service (\udos) attacks
  have been demonstrated to fundamentally bypass
  traditional DoS mitigation schemes
  based on the flow and volume of network packets.
  Their three characteristics---%
  low capacity (volume), slow speed (velocity) and benign looking (legitimacy),
  make it infeasible to
  tame them by simply observing external network activities,
  demanding a finer-grained scheme
  that can monitor the internal activities.
  Recent \udos attacks appear to be not just stealthy
  but also destructive
  so that they often result in severe consequences to the victim machine:
  e.g., SegmentSmack and FragmentSmack attacks
  demonstrated in 2018
  can result in a kernel panic or system hang
  while requiring virtually no resource
  from the attacker.

  In this paper,
  we propose a data-driven approach, called \sys,
  that accurately tracks internal resource utilization
  and allocation associated with each packet (or session),
  making it possible to
  tame resource exhaustion caused by \udos attacks.
  Since \sys focuses on capturing the symptom of DoS,
  it can effectively mitigate
  previously unknown \udos attacks.
  To enable a finer-grain resource tracking,
  \sys provided in concept the accounting capabilities
  to each packet itself, so we called \emph{data-driven}:
  it monitors resource utilization
  at link, network, transport layers in the kernel,
  as well as application layers,
  and attributes back to the associated packet.
  Given the resource usages of each packet,
  \sys can reclaim (or prevent) the system resources
  from malicious packets (or attackers)
  whenever it encounters system-wide resource exhaustion.
  To provide a light-weight resource tracking,
  \sys carefully multiplexes
  hardware performance counters
  whenever necessary.
  Our evaluation shows that
  \sys's approach is indeed effective
  in mitigating real-world \udos attacks
  with negligible performance overheads---%
  incurring 3\%--4\% throughput and latency overheads
  on average when the system is throttled.

\end{abstract}

\section{Introduction}
\label{s:intro}
Denial-of-Service (DoS) attacks are
ongoing, evolving threats to the Internet.
Basically,
they throttle a victim server
with a tremendous number of network packets,
so that the server becomes unresponsive to
other benign requests.
To maximize the number of attack packets,
they often abuse a large number of zombie computers~\cite{zombie-roundup},
also known as distributed DoS (DDoS) attacks.
Existing defenses or mitigation approaches
either rely on a good, reactive infrastructure to handle higher network capacity,
limit the rate of packets to which each client can respond, or
filter out the packets with known bad signatures%
~\cite{cloudflare-ddos, cloudflare-rate-limiting, project-shield}.
These approaches, however,
make DoS attacks evolve into two other extremes:
either maximizing attack volume
to overwhelm the defenses
(e.g., the 1.7~Tbps Memcached reflection attack~\cite{memcached-reflection}),
or minimizing attack volume to bypass them
(e.g., the SegmentSmack~\cite{segmentsmack,segmentsmack-freebsd} and FragmentSmack~\cite{fragmentsmack} attacks).
This paper focuses on the latter attack as known as a \emph{low-volume DoS attack} (\udos).
\udos requires low-volume, slow-rate attack traffics,
yet still results in destructive consequences
to the victim machines.
Instead of merely flooding a victim's network,
\udos attacks send either crafted exploit packets
to trigger DoS vulnerabilities of target software,
or expensive, yet legitimate requests
that can easily exhaust important system resources,
such as CPU cycles, memory or file descriptors.
Conventional DoS defense schemes
such as improving network capacity,
restricting the packet rates or even blocking known bad requests,
are, unfortunately, less effective
against \udos attacks.
For example,
recent \udos attacks,
\segmentsmack~\cite{segmentsmack,segmentsmack-freebsd} and
\fragmentsmack~\cite{fragmentsmack},
rely only on legitimate packets
that can exhaust the CPU cycles of a victim's servers
by stressing the re-assembling logic
that handles out-of-order TCP segments and incomplete IP fragments.
%
%
More traditional \udos attacks,
such as Billion Laugh~\cite{billion-laugh} and Slowloris~\cite{slowloris},
can cause memory exhaustion or limit the number of active sessions
of the victim's server,
while relying only on benign-looking requests
that can bypass the existing DoS defense schemes.

Existing countermeasures against \udos
focus only on fixing specific instances of attacks
since by definition, all \udos attacks
are hardly identifiable
in terms of network traffics and legitimacy.
It is important to address the root cause of \udos attacks,
but our communities are desperately
looking for more universal, practical solutions
that can mitigate the emerging \udos attacks~\cite{darpa-xd3, slowloris}.
To be deployable in practice,
it is also important to
address the \udos problems
at the end points (\ie, hosts)
without requiring modification of network infrastructure.

%
%

In this paper,
we propose \sys, a data-driven resource accounting system
to mitigate \udos attacks
by monitoring how each packet (or session)
accounts to the system resource usages,
and reacting to the packet causing resource exhaustion.
The key idea of \sys is to focus
on identifying the \emph{symptom} of \udos
(\ie, resource exhaustion, so DoS),
rather than identifying specifics of attacks,
such as exploit methods or types of DoS vulnerabilities.
Once \sys concludes that the systems are throttled
or potentially under a DoS attack,
it reacts to the current situation
based on the amount of the system resources
attributed to each packet (or session).
As a reaction,
\sys attempts to reclaim the resources
of the most exhaustive packets (or sessions) in order,
and prevent the future attacks
by blacklisting their origins.

To make this data-driven resource accounting possible,
\sys attempts to accomplish three goals:
precision, performance and generality.
First,
\sys implements a finer-grained resource accounting scheme.
It provides in concept resource accounting capabilities
to each packet (or session);
it monitors resource utilization
at link, network, transport layers in the kernel,
as well as application layers,
and attributes back to the associated packet.
This allows \sys to reason
about \udos attacks
targeting specific network layers.
Second,
\sys minimizes the performance overheads
required for the fine-grained resource accounting
by carefully multiplexing
hardware performance counters (HPCs).
Third,
\sys focuses on identifying the symptom of the DoS attacks
and mitigates them
without knowing the details of the attack methods nor
targeted resources for exhaustion.
Most importantly,
\sys is designed to tackle \udos
at the end hosts
without requiring the modification
of the network infrastructure.

Our evaluation shows that
\sys can identify real-world \udos attacks
(\eg, FragmentSmack, Apache Range Header, and Slowloris)
targeting various types of system resources
(\eg, CPU, memory, and connection pool)
at different layers (\eg, network, transport, and application layers).
Also, \sys can mitigate on-going \udos attacks
by selectively blocking exhaustive requests.
That is,
it can continue to serve legitimate requests
by dropping up to 16\% of requests
even under active \udos attacks.
%
It also imposes negotiable performance overheads:
when the system is throttled and \sys is applied,
it incurs only 3.5\%--4.8\% of latency and throughput overheads.
\vspace{2pt}
The summarized contributions of this paper are as follows:
\squishlist
\item \textbf{Data-oriented resource-usage profiling.}
  To the best of our knowledge,
  \sys is the first study
  that detects and blocks suspicious requests
  according to their high system resource usages.
  \sys enables data-oriented resource-usage profiling to
  accurately identifies how many resources have been used
  to process each request at each network layer,
  detecting suspicious clients
  regardless of which unknown exploit techniques
  they use.
  %
\item \textbf{Hardware-based efficient profiling.}
  \sys uses HPCs for efficient profiling of
  per-packet per-layer resource usage.
  More specifically,
  \sys uses the performance monitoring unit (PMU)
  to check the number of retired CPU instructions at each layer, and
  the memory bandwidth monitoring (MBM)
  to check the number of memory accesses at each layer
  for processing each packet.
\item \textbf{Universal defense.}
  \sys is effective against various \udos attacks
  targeting different resources.
  %
  \sys monitors various system resources simultaneously
  using the same technique
  to detect and avoid any of their exhaustion.
  %
\squishend
The remainder of this paper is organized as follows.
\autoref{s:background}
defines \udos attacks and
explains previous studies.
\autoref{s:motivation} introduces
our motivation, research goal, and challenges.
\autoref{s:design}
depicts the detailed design of \sys.
\autoref{s:impl}
describes how we implemented \sys.
\autoref{s:eval}
explains our case studies on \udos attacks and
evaluates the performance overhead and the effectiveness of blocking in \sys.
\autoref{s:discussions} discusses
some limitations of \sys and
potential solutions.
\autoref{s:relwk} introduces related work and
\autoref{s:conclusion} concludes this paper.

\section{Background}
\label{s:background}
We define a \udos attack and
characterize their properties
by using real-world \udos attacks as examples.
We also explain existing approaches and
their limitations,
highlighting the motivation of \sys's approach.

\subsection{Low-volume DoS (\udos) Attack }
\udos attacks aim to
make victim servers unavailable
with a small number of attack packets.
%
They often rely on
carefully crafted, benign looking packets
to effectively exhaust
the important system resources of victim servers,
by exploiting their performance bugs or
heavy operations.
For example,
the event handler poisoning attack~\cite{redos-study, node-cure}
overloads a single-threaded event-driven server (e.g., Node.js)
by requesting expensive computations
such as evaluating complicated regular expressions.
Second,
to exhaust both memory and CPU of a victim server,
some attacks exploit its performance bugs.
%
The BlackNurse attack~\cite{blacknurse}
exploits a vulnerability of a Linux-based firewall that
consumes many CPU cycles and much memory to process
``Destination Unreachable/Protocol Unreachable'' messages of the ICMP protocol.
%
%

Existing countermeasures against \udos attacks
tend to be attack-specific,
as discussed in \cite{splitstack}.
For example,
to mitigate the SSL/TLS Renegotiation attack,
it is recommended to disable the SSL/TLS Renegotiation protocol~\cite{ssl-tls-renegotiation}.
Similarly,
to address the event handler poisoning attack described above,
typical countermeasures are
to estimate the complexity of each regular expression.
Not surprisingly,
it is recommended to restrict the number of active sessions
a client can initiate
in order to mitigate the session-pool \udos attacks~\cite{redos-study, node-cure, slowloris}.
%
%
We believe countering \udos attacks by exploiting attack-specific
characteristics
falls short in mitigating unforeseen, constant
threats of \udos attacks.

\subsection{Defenses Based on Resource Profiling}
\label{ss:prev-work}
Two recent countermeasures~\cite{rampart, node-cure}
have been demonstrated
to identify suspicious requests for CPU exhaustion
by profiling resource usages.
Rampart~\cite{rampart} measures per-function CPU time
for handling individual requests.
It is effective in detecting anomalous requests
that incur drastically longer execution time
than normal requests.
%
Node.cure~\cite{node-cure} aims to
mitigate event handler poisoning attacks
using timeout exceptions.
Node.cure defines wall-clock timeout values
for specific event handlers,
so
they would not spend more time
to process requests
than the defined timeout value.
%

These two approaches are effective in
detecting and mitigating the described \udos attack scenarios,
but are hardly possible to generalize further
to mitigate other types of \udos attacks,
for the following reasons.
First,
they profile resource usages
either in a too course-grained manner
(\ie, per process)---%
failed to attribute specific types of resource exhaustion,
or only at a too high level
(\ie, application layer)---%
failed to accurately accommodate
system-wide noises
such as context switching and interrupt timing.
%
%
%
Second,
they fail to handle \udos attacks targeting
kernel-level resource exhaustion
such as link, network, and transport layers,
which rapidly become popular in recent years
(\eg, SegmentSmack and FragmentSmack in 2018~\cite{segmentsmack,fragmentsmack}).
%
%
Since commodity operating systems
such as Linux and Windows
are monolithic,
it is challenging to profile each network layer
in a non-intrusive manner.
SplitStack~\cite{splitstack} attempts to overcome this problem
by splitting the network stack layers and
profiling each of them.
%
However, it demands huge kernel modification.
Third,
they focus on a single type of hardware resources (\ie, CPU),
thereby failing to provide an accurate view
to multiple hardware resources as well as
software-abstracted resources such as the connection pool.
Since it is not uncommon to exhaust multiple resources
simultaneously~\cite{apache-killer-article},
we need better a \udos mitigation scheme
that can accurately reason about
a diverse set of resource consumption
together.

\subsection{Hardware Performance Counter (HPC)}
\label{ss:hpc}
HPCs are hardware units
to count low-level events of micro-architecture during runtime~\cite{exploiting-hpc}.
With HPCs,
we can efficiently profile
various micro-architectural events related to
program execution, such as
retired CPU instructions, cache hits or misses, and branch predictions and misses.
The number of HPC registers,
however, is quite small
(e.g., four in Intel CPUs~\cite{intel-manual} and
six in AMD and ARM CPUs~\cite{amd-manual, arm-manual})
such that
careful scheduling is necessary
to fully leverage them to
monitor various events simultaneously.

\section{Motivation and Challenges}
\label{s:motivation}
In this section,
we explain our motivation and research goal,
and the challenges we have to overcome.

\subsection{Motivation and Research Goal}
\label{ss:motivation-goal}
Although \udos attacks are serious security threats,
we still lack an effective and general defense mechanism against them
(\autoref{s:background}).
We believe resource-usage profiling is
a promising direction to cope with \udos attacks
because, regardless of which tricks they exploit,
they eventually aim to exhaust important system resources.
Existing approaches, however,
are not accurate enough to figure out
the direct relationship between each request and resource usage
in different layers for various types of resources
(\autoref{ss:prev-work}).
%



\sys is designed to solve this challenging problem
by enabling \emph{data-driven resource tracking}
to avoid heavy resource usage
rather than attempting to detect attacks.
Since \udos attacks exploit
benign yet expensive operations or unknown performance bugs,
it is almost impossible to figure out their intention.
%
Instead,
\sys identifies which data (i.e., network packets)
lead to heavy resource consumption to
block or postpone processing further packets from the origin.
This protects any type of system resources from potential exhaustion.
From the viewpoint of system resources,
whether a packet intended to attack them
is meaningless because
in either case
they cannot serve the packet and subsequent ones
if they are (almost) saturated.
%
%


\subsection{Challenges}
\label{s:challenges}
Realizing a data-driven resource tracking system, \sys, is challenging,
especially because we aim to make it
highly accurate, efficient, and universal.
We specify the three critical challenges of \sys and
explain how we tackle each one.

\PP{C1. Accurate resource usage tracking at each layer}
%
\sys aims to prevent each packet from
exhausting system resources
at each network stack layer.
This is because
some \udos attacks (e.g., BlackNurse, SegmentSmack,
and FragmentSmack~\cite{blacknurse, segmentsmack, fragmentsmack})
tend to exhaust resources only at a specific layer,
which can be hidden when we solely monitor
system-wide resource usage.
\sys solves this challenge by
probing the resource usage
at the entry functions of each layer
that every packet should go through (\autoref{ss:data-oriented-monitoring}).
%

\PP{C2. Efficient resource usage tracking}
%
Per-packet and per-layer resource usage profiling
can induce significant computation and memory overhead.
%
\sys solves this challenge by using HPCs to
accurately and efficiently monitor CPU and memory usage (\autoref{ss:data-oriented-monitoring}).

\PP{C3. Universal mitigation}
Universal defense mechanisms against \udos attacks
are necessary to deal with variants or unknown attacks.
Fixing individual problems and monitoring specific resources
cannot achieve such goals.
\sys solves this challenge
by monitoring various system resources at different layers
simultaneously
to detect and block suspicious packets with heavy resource consumption.

\section{\sys}
\label{s:design}
\sys is designed with three main design principles:
monitoring resource usage in
a fine-grained (per-packet and per-layer in the kernel as well as the application) and
unified manner
while minimizing performance overhead.
We first explain \sys's threat model and
depict its design in detail.


\PP{Threat model}
\sys aims to mitigate \udos attacks
with low capacity, slow speed, and benign looking.
Considering a large volume of network traffic,
i.e., DDoS,
and signature- and behavior-based filtering
are out of its scope.
Different countermeasures~\cite{ddos-survey,ieee-ddos-survey,ddos-taxonomy}
can be used to deal with them.
\sys differentiates and blocks each suspicious host
according to its IP address,
implying that it might be vulnerable to IP spoofing attacks.
Preventing IP spoofing attacks is
a challenging problem requiring
other advanced mechanisms
(e.g., ingress and egress filtering~\cite{rfc2267-ingress-egress-filtering} or
IP authentication header~\cite{rfc4302-ip-authentication-header}).
We plan to adopt such mechanisms to make \sys be robust against IP spoofing.
Lastly,
since \sys relies on HPCs to profile resource usages,
it requires either bare-metal machines~\cite{aws-bare-metal,oci-bare-metal}
or virtualized performance counters~\cite{ec2-pmc}.

\subsection{Overview}
\begin{figure}[t]
    \centering
    \includegraphics[width=1\columnwidth]{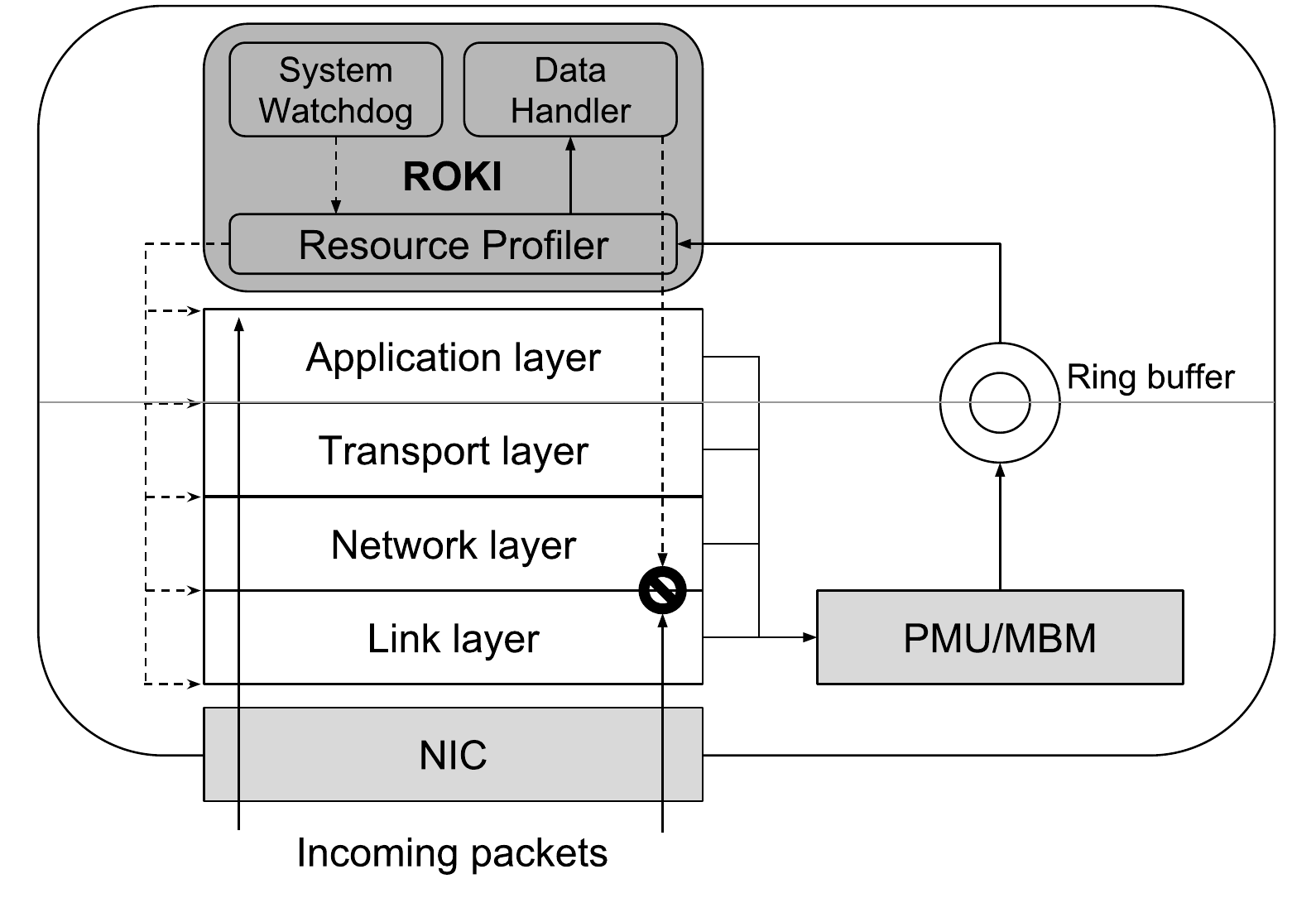}
    \caption{Design overview. \sys consists of resource profiler, system watchdog,
      and data handler. It performs data-oriented resource profiling
      with HPCs while blocking suspicious clients
      based on profiling results.}
    \label{f:design}
\end{figure}

\autoref{f:design} shows a design overview of \sys.
\sys consists of three components:
\emph{resource profiler}, \emph{system watchdog}, and \emph{data handler}.
The resource profiler, the main component of \sys,
monitors the resource usages for handling each packet
at each layer of the network stack, including the
link, network, transport, and application layers.
To achieve this goal,
the resource profiler injects probing code into all of the layers
to keep track of resource usage.
The probing code leverages HPCs to
efficiently profile resource usage,
such as performance monitoring unit (PMU) and
memory bandwidth monitoring (MBM).
%
%
The data handler retrieves per-packet resource-tracking information
for further analysis and resource-exhaustion mitigation,
such as temporal blocks on suspicious clients
that routinely send expensive packets.
%






\subsection{Components}
We explain the three main components of \sys:
resource profiler, system watchdog and data handler.
They run asynchronously to minimize performance degradation.

\PP{Resource profiler}
The resource profiler monitors
the resource consumption of each packet
throughout the system,
following its data flow
from the link layer to the application layer.
To avoid directly modifying the Linux kernel and server applications,
we implemented the resource profiler on top of the bcc framework~\cite{bcc}
along with customization for MBM tracing.
The bcc project seamlessly integrates probing technologies
(e.g., kprobe, uprobe, usdt, and eBPF~\cite{lwn-kprobe, lwn-usdt, cbpf, lwn-ebpf, lwn-ebpf-intro})
into a single framework, which helps us write the tracing code.

During initialization,
the resource profiler populates eBPF code
into probing points located at the entry and exit points of each layer.
The inserted code reads the PMU and MBM counters whenever a
packet hits any of these probing points, calculates the
difference between the two probing points of every protocol layer,
and stores them in a key-value store.  Whenever a packet is
completely processed, its profiling result is
delivered to the resource profiler running
in user space.

\PP{System watchdog}
The system watchdog monitors system-wide resources
to selectively activate blocking.
%
A small Python module periodically checks system-wide CPU and virtual memory usage,
and the number of established connections.
%
Based on the given thresholds for each system-wide resource,
the watchdog activates
the data handler's blocking feature when it observes any suspicious resource usage.
Later, if \sys remedies an attack
such that the resource pressure is relieved, the watchdog deactivates
the blocking function to minimize benign clients who encounter the block.

\PP{Data handler}
The data handler has the two roles: logging and blocking.
Logging allows an administrator to analyze the behaviors of attacks
during real time or off time.
%
Also, the data handler blocks suspicious clients.
Using the mitigation algorithm discussed in \autoref{ss:algorithm}, the data handler
examines clients with corresponding profiling information to decide
what client among them should be blocked.


We used passive and active approaches together to block a certain client. 
Once \sys determines a client as suspicious,
the data handler spawns a thread to block the client asynchronously.
The blocking thread first blocks the suspicious IP address 
with iptables~\cite{iptables} or nftables~\cite{nftables}, and then actively disconnects
established sessions with the blocked IP address
by injecting a shutdown system call into the target process
that manages the sessions via Frida~\cite{frida}.
After the blocking time configured by the administrator passes,
the thread unblocks the client.

\begin{figure}[t]
    \centering
    \includegraphics[width=1\columnwidth]{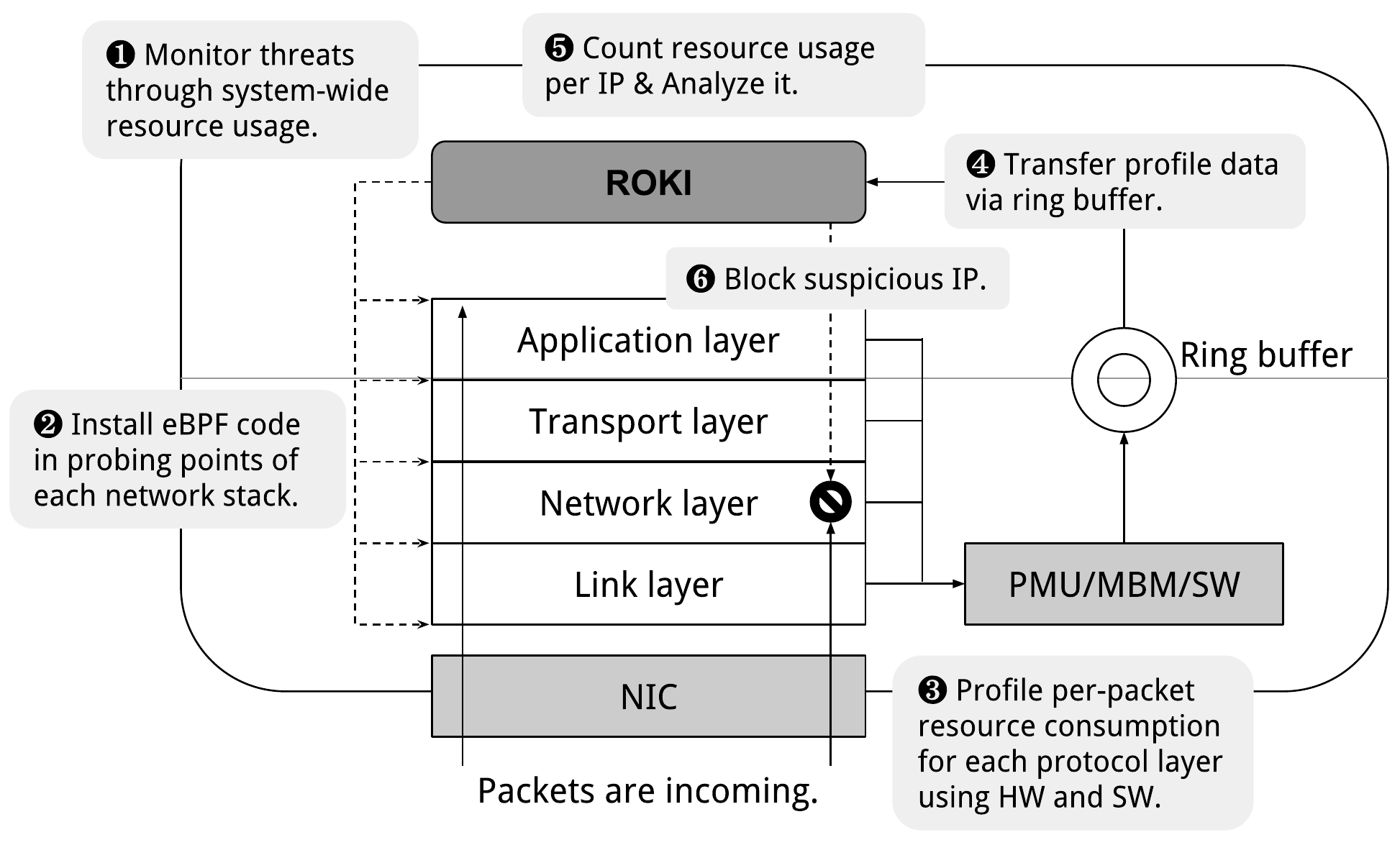}
    \caption{Overall workflow. When \sys begins, it compiles and populates eBPF code
      at the entry points of each network layer, which reads PMU and MBM counters
      whenever packets hit the probing points.
      These values are conveyed via a ring buffer to the user space daemon of \sys.
      Based on these profile results, \sys determines what client is suspicious and blocks it
      for given seconds.}
    \label{f:workflow}
\end{figure}

\subsection{End-to-end Workflow}
\label{ss:workflow}
%
\autoref{f:workflow} shows a detailed workflow of \sys.
First, \sys starts with basic policies including
the default blocking time, inspecting window period, whitelist,
system resource thresholds for turning on and off the resource
profiler, and resource thresholds to restrict the resource usage
of each host.
Then, the system watchdog and the resource profiler start
to monitor system-wide resources and per-packet resource usage, respectively.
If the watchdog detects abnormal activities against the given policies
for those system resources,
it activates the blocking function of the data handler.

\begin{table}[t]
  \centering
  \footnotesize
  \begin{tabular}{ll}
  \toprule
  \textbf{Layer} & \textbf{Entry functions} \\
  \midrule
  \textbf{Application} \\
  ~~Apache  & \cc{ap_invoke_handler}, \cc{ssl_hook_pre_connection} \\
  \textbf{Transport} \\
  ~~TCP  & \cc{tcp_v4_rcv} \\
  ~~UDP  & \cc{udp_rcv} \\
  ~~ICMP & \cc{icmp_rcv} \\
  \textbf{Network} & \cc{ip_rcv} \\
  \textbf{Link}    & \cc{__netif_receive_skb} \\
  \bottomrule
\end{tabular}

  \vspace{1px}
  \caption{Probing points used by \sys. For the transport layer, \sys uses three probing points:
    \cc{tcp_v4_rcv}, \cc{udp_rcv} and \cc{icmp_rcv}.
    \cc{ap_invoke_handler} and \cc{ssl_hook_pre_connection} are
    used by Apache, as probing points for the application layer.}
  \label{t:probes}
  \vspace{-2mm}
\end{table}

The resource profiler monitors the kernel and the target application simultaneously
but asynchronously. More specifically, it consists of two threads that listen to the kernel
and application, respectively.
During initialization,
the resource profiler installs eBPF code
into probing points,
i.e., the entry functions of each layer,
shown in \autoref{t:probes}.

\begin{figure}[t]
  \input{code/ebpf.c}
  \coderule
  \caption{Simplified eBPF code snippet that profiles resource usage at
    the transport layer. It invokes the \cc{perf_read} to retrieve PMU and MBM values.
    To keep consistency, we extended the bcc framework to allow accessing the MBM counter.
    }
  \label{f:ebpf-code}
  \vspace{-17px}
\end{figure}

Once a packet arrives at the server, it is processed throughout
each layer of the network stack in order
where \sys populates eBPF code.
When a packet hits a probing point, the eBPF code resolves the
IP address of the packet first, reads the PMU and MBM counters
afterward, and stores the counts as initial values in a
key-value store at the end.
\autoref{f:ebpf-code} shows the simplified code that performs
this process at the transport layer in the kernel.
If the server has multiple active CPU cores, it needs multiple
key-value stores for each core. The initial value in the
key-value store is used to calculate resource usage for a
single network layer when the processing packet leaves the
current layer or proceeds to the next layer.
Since the CPU core used to process a packet may change,
we also monitor this activity via a separate probe for
\cc{finish_task_switch()}.  When a CPU switch occurs, we calculate
intermediate resource usage on the leaving core, save the
intermediate result in the key-value store, and aggregate
resource usage on the switched core into it later.
Likewise, by installing an additional probe for \cc{inet_csk_accept()},
we monitor the number of connections newly established for a client
while a request sent by the client is processed.
All of the profiling information
will be conveyed to the user-space data handler
via a ring buffer
for further analysis and mitigation.


The data handler groups per-packet profiled data
from the resource profiler based on
the IP address, and holds them in a key-value store.
Meanwhile, it discards stale data in the key-value store
according to configuration.
To decide which client is suspicious, the data handler uses the
algorithm described in \autoref{ss:algorithm}.  With the data received in
the past, the handler ranks IP addresses per resource and
determines which client ranks at the top in terms of each resource usage.
If the sum of resource usage of the top ranked IP violates a
policy that the administrator sets, the data handler concludes
that the top ranked address is suspicious and prevents the
corresponding client from accessing the server for a while.
The data handler uses passive and active approaches
for this temporal blocking:
updating iptables rules to block the suspicious IP address and
forcefully disconnecting already established sessions.

%

\subsection{Data-oriented Resource Usage Profiler}
\label{ss:data-oriented-monitoring}
For each packet entering into a server,
\sys tracks how many resources have been used to process them
by following its data flow within the server
from the link layer to the application layer.
A \naive approach to achieve this goal is
to instrument all functions that receive a network packet as input
to profile resource usage, which
results in too many changes and heavy performance overhead.
Rather,
\sys focuses on the \emph{entry functions} of each layer
(e.g., \cc{ip_rcv}, \cc{tcp_v4_rcv})
that every network packet goes through~\cite{kernel-flow}.
By measuring resource usage at the entry functions of different layers
and comparing them,
\sys is able to determine
how many resources have been utilized
for each packet at each layer.
Also, instead of directly modifying or instrumenting the entry functions,
\sys uses the Linux kernel's tracing functionalities
(e.g., kprobe, uprobe, and eBPF~\cite{lwn-kprobe, cbpf, lwn-ebpf, lwn-ebpf-intro})
to dynamically inject probing code to them (\autoref{s:impl}).
This flexible approach allows much portability
for developers to apply \sys to any distributed system
beyond a single end point machine
by defining data to track across the distributed system.

To accurately and efficiently profile resource usage,
\sys's probing code uses
the PMU to measure how many CPU instructions were retired and
the MBM to measure how much memory was accessed
during processing each packet.

\subsection{Mitigating Resource-exhaustion}
\label{ss:algorithm}
The data handler and resource profiler of \sys work together to
mitigate system resource exhaustion resulting from potential \udos
attacks, by temporarily blacklisting suspicious clients
identified by \sys that heavily occupy system resources.
%
%
%
%
Our approach consists of two main steps:
(1) identifying which client uses resources the most
and (2) determining if its resource usage violates an administrator's policy.
When a packet arrives on the server,
\sys profiles how it affects system resource usage and
maintains this information for each client that has a unique IP
address.
The gathered data is kept for a certain time frame determined by an administrator.
%
Then,
\sys ranks the clients according to how many resources they have used.
We anticipate that
the top-ranked client will continuously spend many resources as usual
such that it is the most beneficial candidate to blacklist.
%

The next step is to determine whether the resource usage of the top-ranked client violates
the given policies by administrators.
Since \sys always monitors the resource usages of each request,
it can calculate average resource usages to
determine reasonable threshold.
Administrators would use static or dynamic threshold to
determine suspicious clients.
For simplify, this paper assumes that they use static threshold.

It is worth noting that
blocking clients based on resource usage is
mainly to keep a server live longer,
not for detecting \udos attacks.
\sys blocks clients only when
the server is almost out of resources,
implying that
the server eventually fails to serve further requests
from other clients.
In addition,
those temporarily blocked clients
can always retry the failed requests later
when the server is no longer busy.
%



\section{Implementation}
\label{s:impl}

We implemented \sys for securing the Apache (v2.2.13 and v2.4.18)
running in a Linux machine (Fedora 27 powered by kernel version 4.13.9).
%
%
%
Note that we consider
multiple versions of Apache
to reproduce \udos attacks
with their original targets.
\sys is implemented with 2,522 lines of Python code and
2,953 lines of C-like eBPF code.
%
%
We will open source the entire code of \sys.
%

\section{Evaluation}
\label{s:eval}

In this section, we evaluate \sys to answer the following three questions:
\squishlist
\item \textbf{Attack detection:}
  How effective is \sys in detecting real-world \udos attacks
  targeting CPU (\autoref{sss:fragmentsmack}),
  memory (\autoref{sss:apache-range}),
  and connection pool (\autoref{sss:slowloris})?
\item \textbf{Quality of Service (QoS):}
  How effective is \sys in maintaining latency? (\autoref{ss:qos})
\item \textbf{Performance overhead:}
  How much performance overhead does \sys incur to profile packets? (\autoref{ss:overhead})
\squishend

\PP{Experimental setup}
We evaluated \sys in a 1GbE local network that consists of four
machines, acting as server, attacker, benign client, and latency
monitor.
The victim server protected by \sys
had two Intel Xeon E5-2687W v4 CPUs (24 cores, 3GHz)
and 252 GB of memory.
The server ran Linux kernel version 4.13.9.
In our evaluation,
we intentionally enabled only four cores of the server to easily
exhaust it unless otherwise stated.
The attacker machine
had an Intel Xeon CPU E7-4820 CPU (16 cores, 2GHz)
and 125 GB of memory.
The benign client machine
was equipped with an Intel Core i7-6600U CPU (4 cores, 2.6GHz)
and 19 GB of memory.
The latency monitor machine
had an Intel Xeon CPU E7-4820 CPU (16 cores, 2GHz)
and 252 GB of memory.
To check response time,
the monitor periodically
sent the server an ApacheBench (ab) request every 0.5 seconds and
calculated the round-trip time.
%


\PP{Default configuration}
\sys is fully configurable for administrators in terms of its
blocking period, windowing time for inspection, system resource
monitoring period for the watchdog, and thresholds for each
network layer. We summarize the configuration used for
our evaluation in \autoref{t:config}.

\begin{table}[t]
  \vspace{0.7em}
  \centering
  \footnotesize
  \begin{tabular}{ll}
  \toprule
  \textbf{Configuration} & \textbf{Default value} \\
  \midrule
  Blocking time  & 5 seconds \\
  Windowing time for inspection & 3 seconds \\
  Monitoring interval for system resource & 0.1 seconds \\
  Ring buffer size per core & 16 MB \\
  \textbf{Conditions to enable/disable Resource Profiler} \\
  ~~CPU usage  & 75\% / 35\%  \\
  ~~Memory usage  & 75\% / 50\% \\
  ~~Connection pool & 75\% / 35\% \\
  \textbf{Instruction Thresholds} \\
  ~~Application  & 300,000 \\
  ~~Transport & 45,000,000,000 \\
  ~~Network & 1,000,000,000 \\
  ~~Link & 80,000,000,000 \\
  \textbf{MBM Thresholds} \\
  ~~Application  & 1,000,000,000 \\
  ~~Transport & 50,000,000,000 \\
  ~~Network &  500,000,000 \\
  ~~Link & 1,500,000,000,000 \\
  \textbf{Connection Threshold} & 6 \\
  \bottomrule
\end{tabular}

  \caption{Default configuration used for our evaluation.}
  \label{t:config}
\end{table}

\begin{figure}[t]
\centering
    \includegraphics[width=0.48\textwidth]{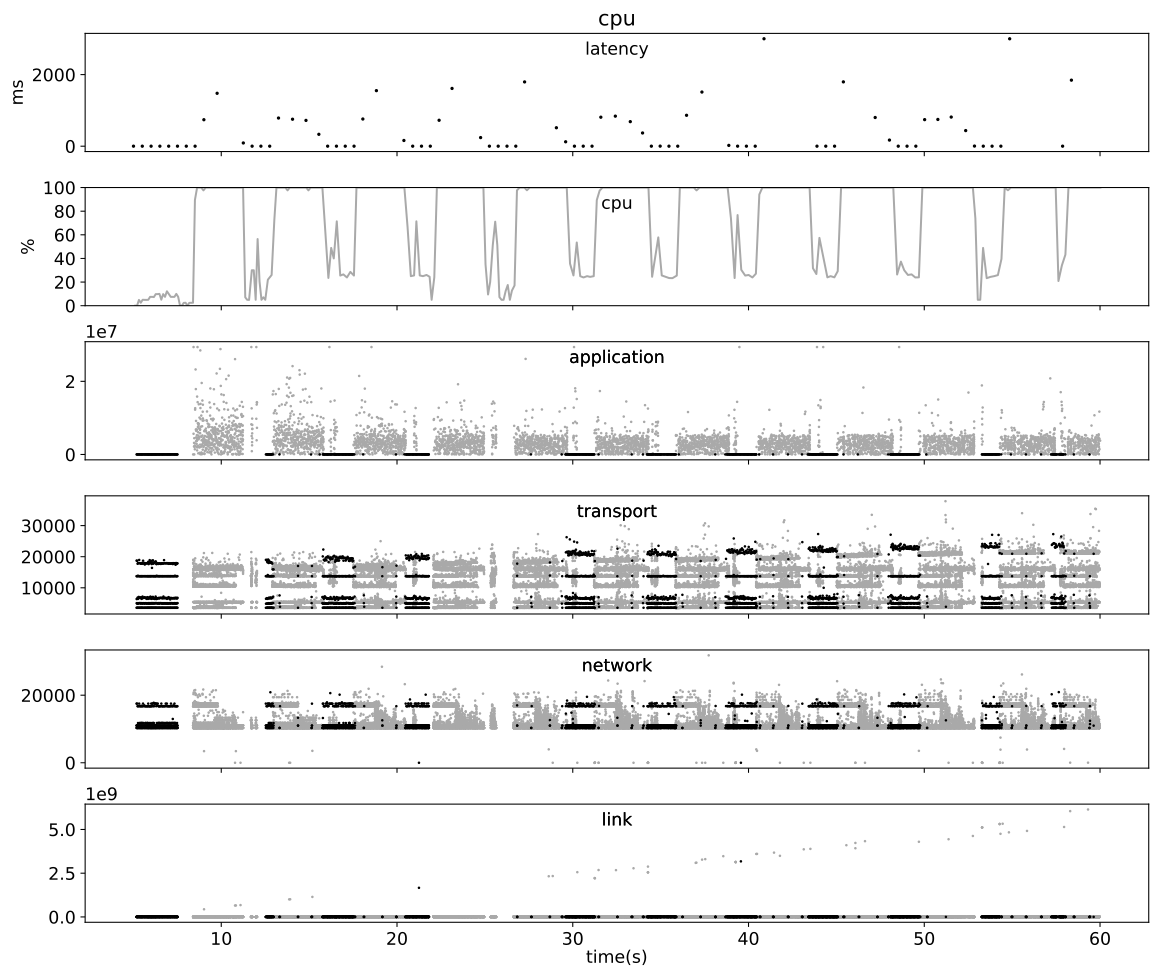}
\caption{Apache Range Header attack profiled by \sys without intervention.
  Gray and black dots represent
  the number of retired CPU instructions for handling each packet
  from the attacker and benign client at each layer, respectively.
  The server's latency, system CPU usage, and number of instructions
  used in the four network layers
  are displayed in order from the top-most plot.}
\label{f:byterange-cpu-pristine-img}
\end{figure}

\PP{Interpretation of experimental results}
Before diving into each experiment in detail,
we explain how to interpret the profiling results from \sys.
\autoref{f:byterange-cpu-pristine-img} shows the CPU utilization of
the server (i.e., the number of retired CPU instructions) for
handing individual network packets at each network layer under
Apache Range Header attacks profiled by \sys.
We monitored the server for 60 seconds where the \udos
attack was conducted.
%
%
Here,
gray dots represent the packets from the attacker machine and
black dots represent the packets from the benign client machine
and the latency monitor.
The top-most plot, which indicates the latency measured
at the latency monitor, is provided for reference and not used
by \sys for detection and mitigation.
%
%
When \sys was fully initialized,
the benign client started sending requests to the server from the 5.5th
second (black dots) and
the attacker started the attack from the 8.5th second (gray dots)
and repeated it every 4.5 seconds. Each attack lasted for 3 seconds.
During the attack,
the server could not handle benign requests such that its latency increased.
However, in respite from the attack, it restarted serving benign requests.

By carefully analyzing the result,
we found that
the server utilized more CPU resources for handling
the packets from the Apache Range Header attack than
those for handing benign packets at the application layer.
%
Thus, by blocking the clients sending expensive packets,
\sys has a chance to mitigate this \udos attack.
%


\begin{figure}[t]
  \centering
  \includegraphics[width=0.48\textwidth]{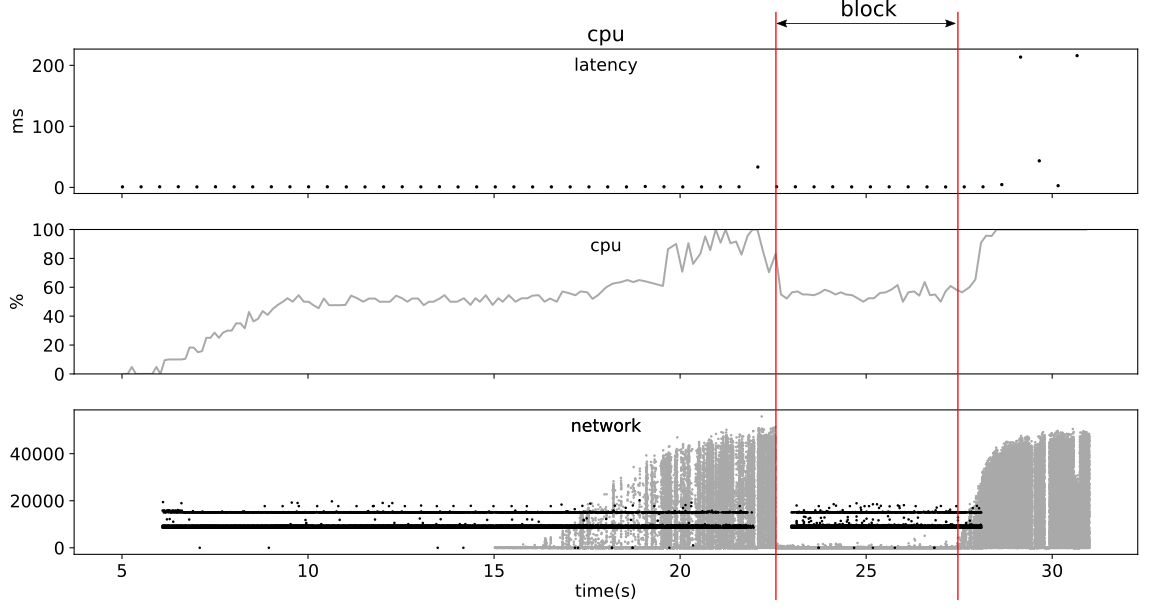}
    \caption{\fragmentsmack under \sys's protection.
      \fragmentsmack attacks the IP fragment reassembly at the network layer.
      \sys detects its heavy resource usage at the network layer and blocks it.}
  \label{f:fragmentsmack-img}
\end{figure}

\subsection{Mitigating Real-world \udos Attacks}
To know the effectiveness of \sys,
we reproduce three representative real-world attacks,
\fragmentsmack (CPU),
Apache Range Header (memory), and
Slowloris (connection pool),
against a server protected by \sys.

\subsubsection{FragmentSmack}
\label{sss:fragmentsmack}
\fragmentsmack~\cite{fragmentsmack}
is a CPU exhaustion attack
that exploits a performance bug
in reassembling IP fragments at the network layer.
In the Internet,
senders can break large packets
(i.e., bigger than the maximum transmission unit)
into a number of fragments,
which will be reassembled by receivers
in regular sequence.
Until the entire fragments arrive,
receivers have to queue arrived fragments
while arranging them in order according to their offset values.
\fragmentsmack aims to prolong this process
by creating and transmitting small fragments with arbitrary offsets
while discarding last fragments,
whose value of the more fragment (MF) bit is zero,
to prohibit complete packet reconstruction.
%


\PP{Methodology}
We reproduced \fragmentsmack according to
\cite{fragmentsmack, redhat-fragmentsmack, ms-fragmentsmack, remco-fragmentsmack}.
With Scapy~\cite{scapy},
we created 64\KB UDP packets and
broke them into 8\B fragments
with the same ID but arbitrary offsets.
Then,
we used 64 workers to transmit the fragments
to a victim concurrently
in arbitrary order
while discarding fragments with MF=0.
%
Single IP \fragmentsmack saturates one CPU core~\cite{redhat-fragmentsmack},
so, in this experiment,
we activated two cores of the victim server for resource saturation.
We increased the high and low threshold for IP fragmentation,
\cc{net.ipv4.ipfrag_high_thresh} and \cc{net.ipv4.ipfrag_low_thresh},
to 64\MB and 48\MB, respectively, at the victim server
to make it have a queue
long enough to incur busy defragmentation.
Instead of iptables that does not handle fragments,
we used nftables~\cite{nftables}
to block the source IP addresses of suspicious fragments.

\PP{Experimental results}
\autoref{f:fragmentsmack-img} shows
how \sys mitigated FragmentSmack.
The victim server received benign requests from the 6th second
and malicious requests from the 15th second.
As fragments from the attacker were queued, we observed that
the victim's CPU usage increased.
\sys blocked the origin of these fragments at the 25th second
because they consumed CPU beyond the given threshold at the network layer.
%
During the blocking period,
the victim server was able to serve benign requests
because
nftables discarded malicious fragments
as soon as they arrived.

\subsubsection{Apache Range Header attack}
\label{sss:apache-range}
The Apache Range Header attack~\cite{apache-range-header-attack}
is a \udos attack that overloads
the CPU and memory of a victim server by exploiting a protocol
design flaw in the Apache web server. More specifically, the HTTP
protocol allows a client to request multiple overlapped ranges
in a single request, which makes the server perform large
fetches that are inefficiently kept in
memory.
Although the Apache
Range Header attack exhausts both CPU and memory, we focus
on memory implication of the attack in our evaluation.

\PP{Methodology}
We used the Apache Killer script~\cite{apache-killer}
against httpd-2.2.13
which is vulnerable to the Apache Range Header attack.
%
For this experiment,
we used ten-second blocking periods to
clearly see \sys's effectiveness.

\begin{figure}[t]
  \centering
  \includegraphics[width=0.48\textwidth]{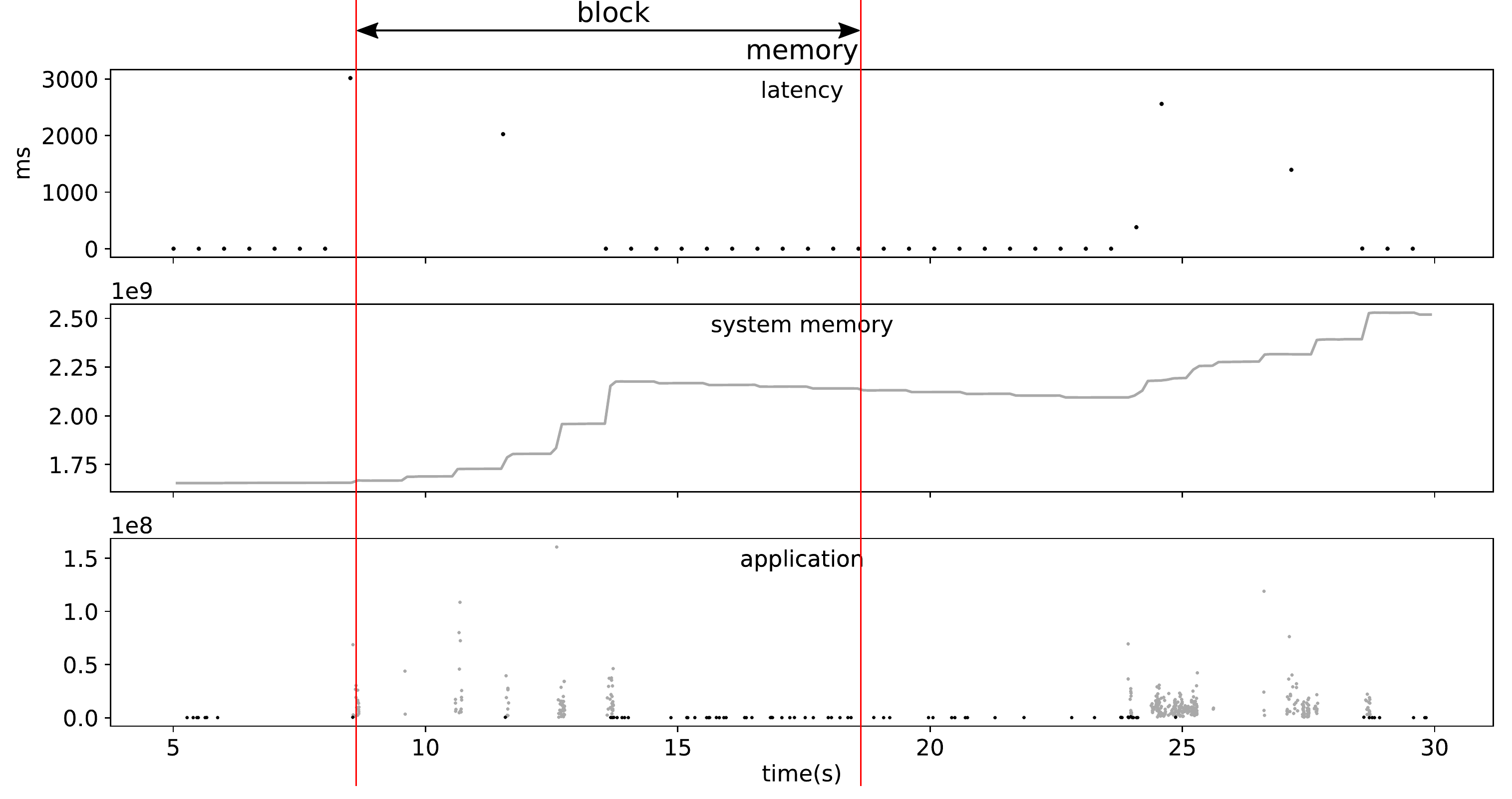}
  \caption{The Apache Range Header attack exhausting memory in the application layer.
    \sys prevented this attack from exhausting the system memory.}
  \label{f:byterange-roki-img}
\end{figure}

\begin{figure}[t]
  \centering
  \includegraphics[width=0.48\textwidth]{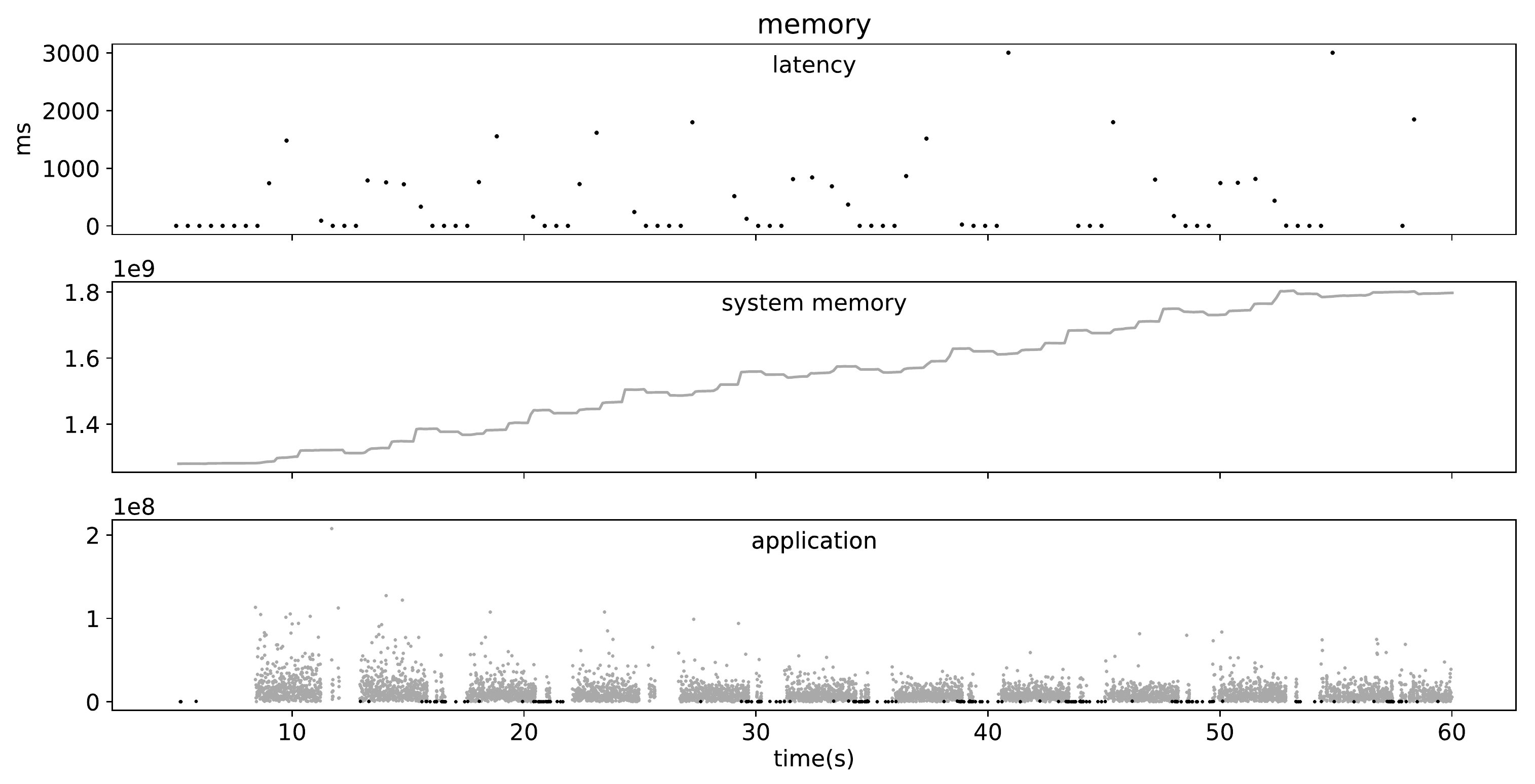}
  \caption{The Apache Range Header attack profiled by \sys without blocking.
    The system memory usage kept increasing due to the attack.}
  \label{f:byterange-pristine-img}
\end{figure}

\PP{Experimental results}
\autoref{f:byterange-roki-img} shows how \sys
mitigated the Apache Range Header attack.
When malicious requests arrived in the server at the 8th
second, \sys successfully detected the abnormal memory movement
in the application layer via high MBM values even before
the overall system memory usage increased.
The suspicious IP was blocked at the 8th second, but the active
memory of the system increased during the blocking
period. This was because the server had to process
arrived requests before the IP was blocked.
However, the active memory steadily decreased later
(i.e., after the 14th second) until the next attack arrived.
Without \sys,
the overall memory usage kept increasing
(\autoref{f:byterange-pristine-img}).

\subsubsection{Slowloris}
\label{sss:slowloris}
The Slowloris attack~\cite{slowloris}
aims to occupy as many connections as possible
to prohibit establishing further benign connections,
by sending partial requests that do not complete.

\PP{Methodology}
To reproduce Slowloris,
we applied the Slowloris script~\cite{slowloris-tool}
against httpd-2.4.18.
We used netstat~\cite{netstat}
for checking the total number of established connections.
We also counted the number of new connections for each client,
as explained in \autoref{ss:workflow}.

\begin{figure}[t]
  \centering
  \includegraphics[width=0.48\textwidth]{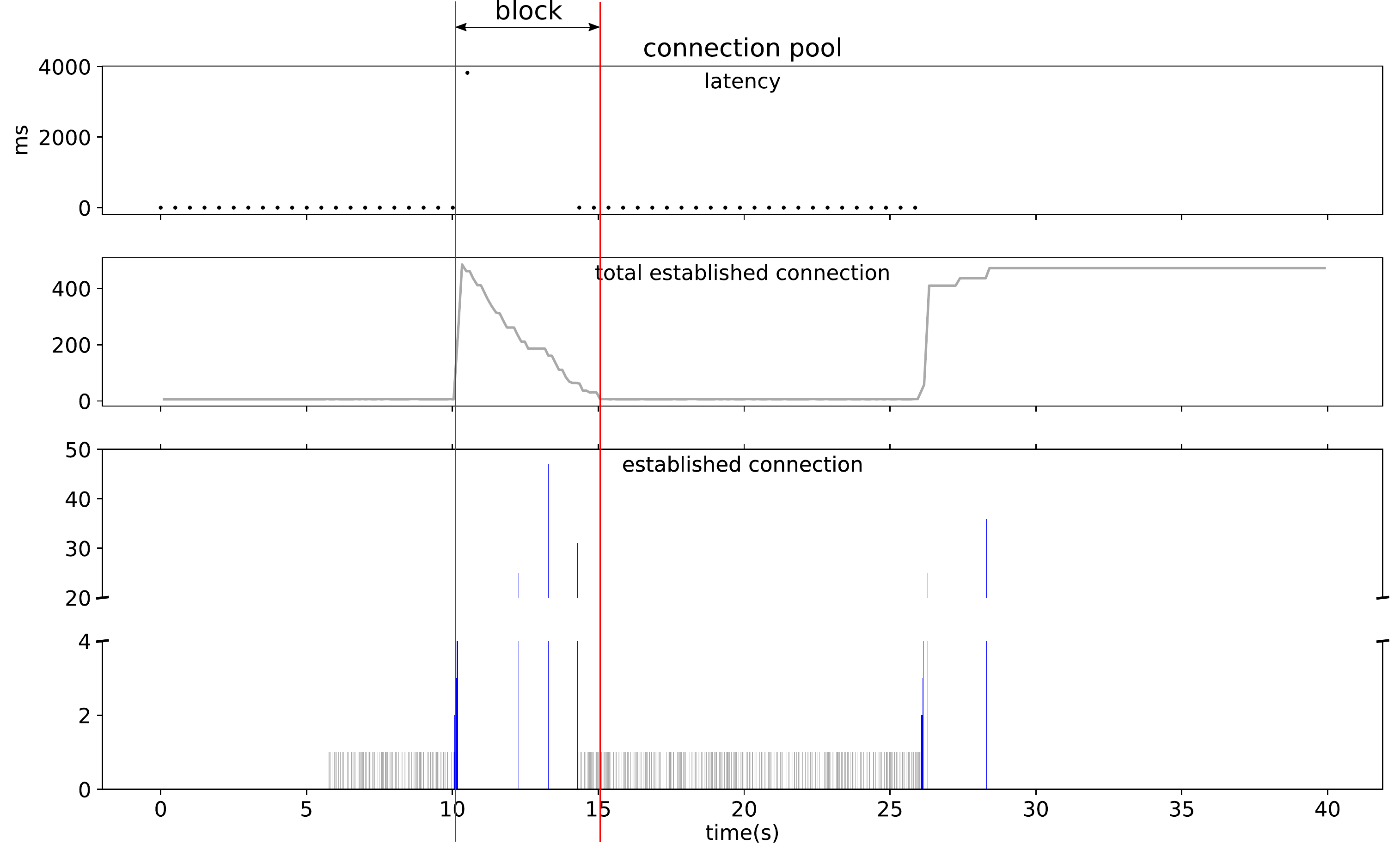}
  \caption{\sys detected Slowloris by counting newly established connections in real time.
    Gray and blue bars represent
    the number of established connections
    with a benign client and an attacker, respectively.
    After \sys blocked the attack at the 10th second, the connections from the attacker
    were closed such that the benign client could make new connections from the 14th second.}
  \label{f:slowloris-roki-img}
\end{figure}

\PP{Experimental results}
\autoref{f:slowloris-roki-img} shows how \sys mitigates the Slowloris attack.
The benign requests (gray bars) that appeared since the 6th
second consistently made a single connection.
In contrast,
the malicious requests (blue bars) that
appeared since the 10th second abnormally established many
connections in a very short time.
This difference allowed \sys to
detect the malicious requests.
Once the attacker was identified, \sys actively withdrew
suspicious connections and injected a shutdown system call into
web server processes via Frida~\cite{frida}. The number of
established connections kept decreasing from the 10th second to the 15th
second, and \sys served the benign client from the 14th second
until the next attack occurred at the 26th second.

\subsection{Latency versus Failure Rate}
\label{ss:qos}
\begin{figure}[t]
  \centering
  \includegraphics[width=\columnwidth]{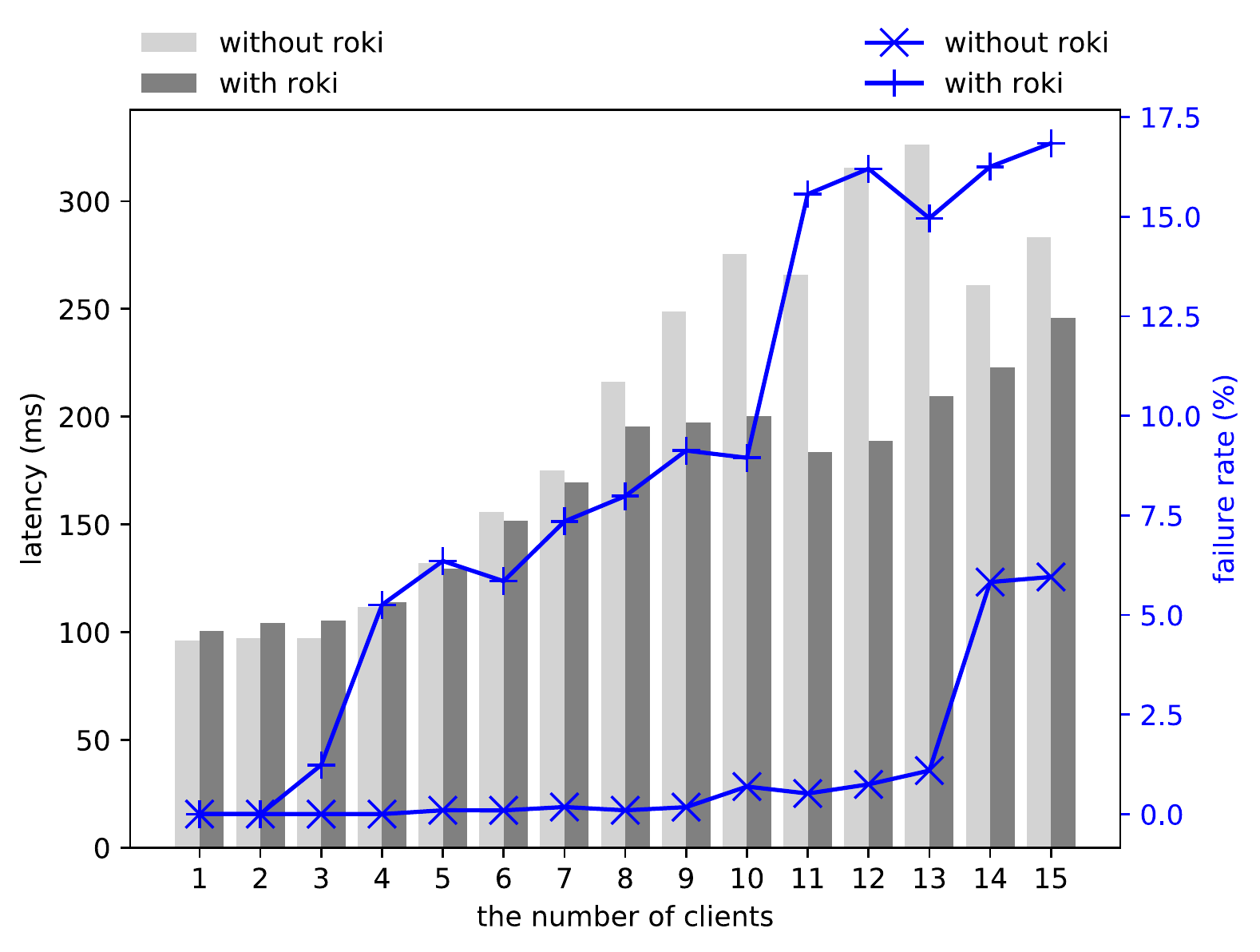}
  \caption{Average latency versus request failure rate
    with the increasing number of concurrent clients.
    \sys maintains reasonable latency for clients
    even when a server is throttled.
    Whenever the system is under exhaustive resource uses,
    \sys drops the most expensive requests at the moment,
    which indeed help system's resource allocation
    and so the QoS of the service.}
  \label{f:qos}
\end{figure}

\sys maintains good QoS even when a server is fully saturated
with \udos attacks or other benign but expensive requests
by selectively blocking clients who consume the most resources.
To verify whether \sys satisfies this goal,
we conducted an experiment to
see how the average latency and
failure rate (i.e., request drop rate)
vary according to the load of a web server
protected by \sys.
%
The server was serving mirrored Wikipedia pages~\cite{wikipedia-mirroring},
and multiple clients from 1 to 15
concurrently requested random Wikipedia pages from the server
for 30 seconds.
We emphasize that we set no attacker in this experiment because
\udos attacks can be constituted with legitimately formatted packets
so differentiating attacker's packets from benign ones is not \sys's main concern.
We measured
(1) the average latency only for
\emph{successful requests} (i.e., requests not dropped by \sys) and
(2) the failure rate (i.e., requests dropped by \sys).
\autoref{f:qos} shows the results.
\sys started to drop some requests
when the number of concurrent clients
was larger than 2.
Without \sys,
the average latency of the server
sharply increased in proportion to the number of concurrent clients.
However,
with \sys,
the slope was gentle with some failure rates.
When the number of concurrent clients was 12,
\sys improved the average latency by up to 1.67\X
while dropping 16.2\% of requests (122 out of 753 requests).
This shows \sys maintains QoS well
by dropping some requests as expected.
%
%
Interestingly,
when the number of concurrent clients was
larger than or equal to 14,
we observed failed requests even without \sys
due to server saturation,
improving the latency.
These arbitrary packet drops,
however,
are problematic because, usually,
there are more normal clients than attackers (or heavy clients)
such that benign clients would observe more failures
compared with the others.
%
%
%

\subsection{Profiling Overheads}
\label{ss:overhead}

\sys's fine-grained resource accounting
unavoidably demands additional CPU and memory resources,
resulting in degraded latency and throughput, and
more memory consumption.
In this section,
we evaluate the profiling overheads
by using Apache Benchmark and
calculating additional memory requirements.

\PP{Latency and throughput}
\label{sss:latency-throughput-methodology}
We measured how \sys affected the latency and throughput of a
web server by running http-2.4.33 on the server machine
protected by \sys and Apache Benchmark (ab) through a client
machine.
The web server enabled 48 cores and
was serving a static page.
ab kept requesting the static page
while varying concurrency from 10 to 100,
and 512 that was the maximum number of sessions
the web server supported.
%
\sys does not block any connections
for this microbenchmark to measure its pure profiling overhead.
We repeated each experiment 10 times and averaged the results.
\begin{figure}[t]
    \centering
    \includegraphics[width=1.0\columnwidth]{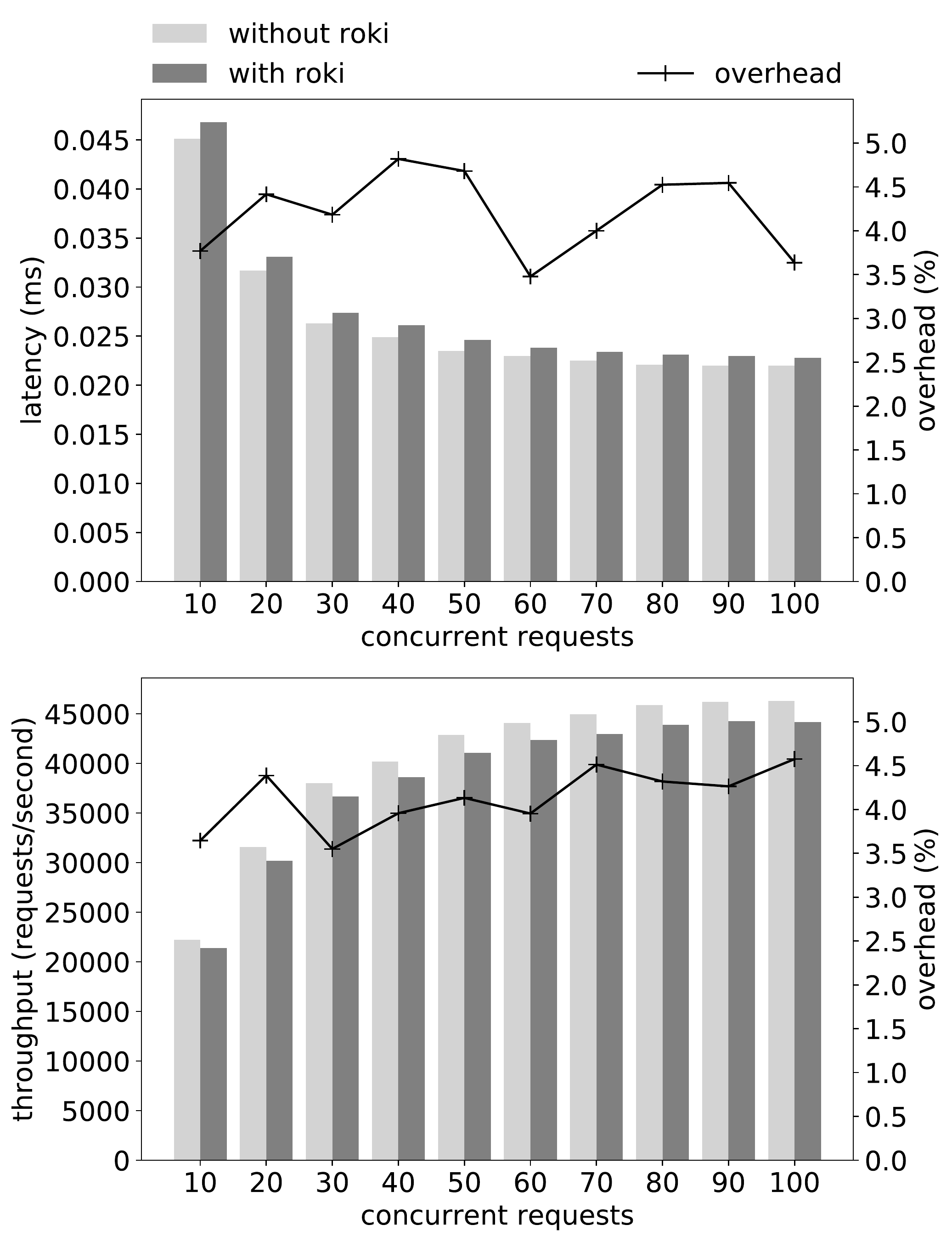}
    \caption{Apache benchmark result with \sys.
    The top figure shows the overhead caused by \sys
    in latency and the bottom one shows the overhead in throughput.
    Light gray and dark gray bars represent results without and with \sys, respectively.
    \sys introduced 3.48\%--4.82\% overhead in latency and
    3.55\%--4.58\% overhead in throughput while the
    number of concurrent connections was varied from 10 to 100.}
    \label{f:overhead-ab}
\end{figure}
%
\autoref{f:overhead-ab} shows the latency and throughput of the server
with and without \sys
whose overhead stably moved between 3\%--5\% according to the number of concurrent connections.
The minimum overhead in latency was 3.48\%
when the number of connections were 60, and
the one in throughput was 3.55\%
when the concurrent connections were 30.
%
%
%
In the worst case
where the number of concurrent requests was 512,
the overhead in latency was 5.98\% and
the overhead in throughput was 4.19\%.

\PP{Memory overheads}
\label{sss:memoverhead}
For each IPv6 packet,
\sys requires 126~bytes
of per-packet profiling data;
meaning that
for queuing 100,000 packets for analysis,
it requires around 12~MB of additional memory uses.
Currently,
\sys uses a 16~MB ring buffer for each core
to satisfy this requirement.
The per-packet profiling data of kernel space
consists of
timestamp (8~B), CPU ID (4~B), source IP address (4~or~16~B),
instruction counters for three layers (3~\X~8~B),
MBM counters for three layers (3~\X~8~B),
and that of userspace has
one more field to store program ID (4~B).

\section{Discussion}
\label{s:discussions}

In this section,
we discuss some limitations of \sys and its optimal configuration.

\PP{Ring buffer overflow}
\sys is not able to comprehensively
analyze or block suspicious packets
if a ring buffer overflows.
\sys uses a ring buffer to deliver
per-packet profiling information collected
at each layer of the in-kernel network stack
to the user-space resource profiler.
Since the ring buffer is finite,
it can overflow if
too many concurrent packets are delivered or
the resource profiler does not efficiently
consume profiling information,
resulting in data loss.
Thus,
an administrator needs to configure
the ring buffer's size sufficiently to
not suffer from buffer overflow
unless there is exceptional network traffic
(likely due to a DDoS attack).

As explained in \autoref{sss:memoverhead},
the current ring buffer size \sys is using,
16~MB is enough to handle normal HTTP requests.
We also want to emphasize that
the ring buffer overflow does not
hurt the server's normal operations.
This is because it only maintains
additional profiling information used by \sys
not by the kernel or server application.

\PP{HPC scalability}
The HPC has a limitation in scalability:
that is,
the number of performance events
it can concurrently monitor is limited.
For example,
the Intel Xeon E5-2687W v4 CPU that
we use for evaluation supports
three fixed-function performance counters and
four programmable counters per logical core~\cite{intel-manual}.
%
%
Currently,
\sys only uses two programmable counters to monitor
retired CPU instructions and memory bandwidth,
so it does not suffer from a scalability problem.
If we want to monitor other performance events as well
(e.g., cache misses and branch instructions)
to detect other types of resource exhaustion,
we need to rely on the kernel's time multiplexing of HPCs.

\PP{Optimal policy}
Figuring out the optimal policy
to determine when to block
what clients is important.
However,
the optimal policy heavily depends on
server configurations and
administrators' performance goal.
Without real environment setup and data,
determining the policy makes no sense.
Our best-effort strategy
to determine threshold values
was calculating average response time
while varying request drop rates
to find acceptable balance between them.
We do not claim that it was the best approach,
but we think that it is one of the feasible approaches
to determine acceptable threshold.


\section{Related Work}
\label{s:relwk}
In \autoref{ss:prev-work}, we discuss
Rampart~\cite{rampart}, Node.cure~\cite{node-cure}, and SplitStack~\cite{splitstack}
to detect or mitigate \udos attacks.
Apart from these strategies, other studies
inspire us to design and implement \sys.
We describe them in the rest of this section.

\PP{Tracing based profiling techniques}
To detect abnormality in resource usage,
\sys relies on a data-oriented resource profiling technique
that records resource consumption, tracing unique data.
Other researchers also consider
how to profile a system
based on tracing techniques for different goals.
Magpie~\cite{magpie} models CPU workload in distributed system,
tracing events incurred by a request.
After correlating OS level events,
Magpie can figure out CPU usage per request.
However, unlike \sys, it is based on event tracing, and
relatively coarse-grained (i.e., no per-layer resource profiling).
Also, it does not profile memory usage.
Pip~\cite{pip} proposed a bug finding technique for applications
running on distributed system
by comparing actual behavior and expected behavior.
To check application behavior,
Pip records paths by tracking explicit path identifiers.
Unfortunately, Pip is not helpful to defeat \udos attacks
whose behavior is almost legitimate.
A general tracing framework was introduced by X-Trace~\cite{xtrace}.
It enables reconstruction of user's task tree and
a comprehensive view across layers and applications.
However, it does not deal with a resource accounting technique
to protect the system from \udos attacks as \sys does.
Besides, it requires source code change to embed metadata
into target software for tracking.
Retro~\cite{retro} presents a resource management framework for a distributed system
of which resource is shared by multiple tenants.
It aims to achieve desired performance guarantee or fairness
for each tenant through profiling per-workflow resource load.
Retro's granularity for resource profiling is coarser than \sys's one, and
its per-workflow resource profiling does not work for \udos attacks
since the attack tends to exhaust resources on an end-point machine.
Monitoring resources across distributed machines hinders
the attack from being detected.

\PP{DoS/DDoS defenses}
DoS and DDoS attacks aim to
send a large number of packets
to a victim server.
Especially,
DDoS relies on a large number of zombie machines (or bots)
to generate an excessive volume of network traffic.
Since they are old and popular attacks,
many researchers have already analyzed them and
proposed effective countermeasures~\cite{ddos-survey, ieee-ddos-survey, ddos-taxonomy}.
For example,
researchers extract statistical patterns from
the massive attack traffic to generate filtering rules.
Also,
they analyze command-and-control traffic
between bots and their masters
to take down the botnet.
To defend against DDoS attacks, a large body of research explored
monitoring and mitigation schemes based on packet signatures.
Such schemes include Randomize-Then-Optimize
randomization~\cite{yang2004defense}, detecting period
pulse~\cite{sun2004defending},
spectral analysis~\cite{chen2006collaborative},
and
modeling~\cite{kuzmanovic:low-rate-tcp, smith:backtracking}.
%
Unlike schemes that are designed for specific
protocols or rely on packet signatures,
\sys detects resource
exhaustion using data collected from system performance monitors.
%
%
DDoS defense mechanisms can complement \sys to make a server secure
against both DDoS and \udos attacks.
%


\PP{\udos attacks and defenses}
Pioneering research on \udos
attacks~\cite{kuzmanovic:low-rate-tcp} demonstrated that a
vulnerability in the TCP timeout mechanism can be exploited with
periodic, short-lived, low-volume traffic.
Such attacks can be extended to Pulsing DoS (PDoS)
attacks~\cite{luo2005new}, which exploit the
Adaptive-Increase-Multiplicative-Decrease (AIMD) algorithm
implemented in the TCP protocol.
The Boarder Gateway Protocol (BGP), which is used to perform
routing sessions on commercial routers, was shown to be vulnerable to
\udos attacks~\cite{zhang:low-rate-tcp-routing}.
\udos attacks have been further generalized
as Reduction of Quality (RoQ) attacks~\cite{guirguis2005reduction},
which cause a system to perform below capacity.

Initial work on \udos attacks proposed two approaches:
router-assisted and end-point min-Retransmission Time Out
randomization~\cite{kuzmanovic:low-rate-tcp}.
However, experiments performed in the initial work on \udos showed
that by limiting the peak rate and burst length of an attack, the
proposed attack could still severely degrade throughput without being
detected by the popular DoS detection algorithm
Random Early Detection, Preferential Drop (RED-PD)~\cite{kuzmanovic:low-rate-tcp}.
%

Vanguard~\cite{luo2006vanguard} detects \udos
attacks by monitoring anomalies in network events, such as abnormal
traffic of outgoing TCP ACK signals or an imbalance of incoming and
outgoing ACK signals.
\sys mitigates a wider range of attacks than Vanguard, with
low overhead by comprehensively monitoring resource
exhaustion using HPCs.

\PP{Hardware-based resource monitoring}
Researchers started to use HPCs for security applications.
SlowFuzz~\cite{slowfuzz} aims to
automatically detect the most expensive inputs
for diverse, well-known algorithms
by continuously measuring their resource usage with HPCs
in a domain-independent manner.
That is,
it is a proactive approach to find performance bugs of a program,
which can complement \sys.

Several malware studies~\cite{hpc-malware-adaptive,online-malware-hpc, hpc-sherlock}
use HPCs to check low-level, accurate behaviors of malware.
They monitor a process with HPCs to determine
whether its resource usage behaviors
are similar to the behaviors of known malware.
This line of research, however, needs to be improved
because
it is difficult to associate low-level HPC values with
high-level user intention~\cite{hpc-myth}.
In contrast,
\sys does not suffer from this challenging problem
because its goal is
to determine which packet consumes many system resources,
rather than
to infer some intention from the packet's resource usage pattern.
\vspace{-13px}

\section{Conclusion}
\label{s:conclusion}
%
A \udos attack is challenging to defeat
because of its low capacity, low speed, and legitimacy.
\sys protects an end-point server from
this sophisticated attack with data-oriented resource usage tracking.
It accurately monitors resource
usages along the data flow per request,
recognizes resource usage anomalies
due to the request, and
temporarily block the request origin
in a unified manner.
Evaluation shows that \sys is effective against
real-world \udos attacks targeting CPU, memory, and connection
pool in either kernel or user space
with acceptable overhead.
%
%



\bibliographystyle{plain}
\bibliography{roki,sslab,conf}

\end{document}